\documentclass{emulateapj}
\usepackage{graphicx}
\usepackage[usenames,dvips]{color}
\def\arcsec{$\,^{\prime\prime}$}

\def\Chandra{${\it Chandra}$}
\def\Suzaku{${\it Suzaku}$}

\newcommand{\Msun}{\ifmmode {M_{\odot}}\else${M_{\odot}}$\fi}
\def\XMMU{XMMU J174445.5-295044}
\def\IGR{IGR J17407-2808}
\def\CXOU{CXOU J174042.0-280724}

\shorttitle{Two Fast X-ray Transients}
\shortauthors{Heinke et al.}

\begin{document}
\title{Two Rapidly Variable Galactic X-ray Transients Observed with \Chandra, XMM and \Suzaku}

\author{C.~O. Heinke\altaffilmark{1}, J.~A. Tomsick\altaffilmark{2}, F. Yusef-Zadeh\altaffilmark{3}, J.~E. Grindlay\altaffilmark{4}}

\altaffiltext{1}{University of Alberta, Dept. of Physics, Room \#238 CEB, 11322-89 Avenue, Edmonton AB, T6G 2G7, Canada; cheinke@phys.ualberta.ca}

\altaffiltext{2}{Space Sciences Laboratory, 7 Gauss Way, University of California, Berkeley, CA 94720-7450, USA}

\altaffiltext{3}{Northwestern University, Dept. of Physics \&
  Astronomy, 2145 Sheridan Rd., Evanston, IL 60208, USA}

\altaffiltext{4}{Harvard University, Dept. of Astronomy, 60 Garden Street, Cambridge, MA 02138, USA}

\begin{abstract}

We have identified two moderately bright, rapidly variable transients in new and archival X-ray data near the Galactic center.   
Both objects show strong, flaring variability on timescales of tens to thousands of seconds, evidence of $N_H$ variability, and hard spectra. 
 \XMMU\ is seen at 2-10 keV fluxes of $3\times10^{-11}$ to $<10^{-12}$ ergs cm$^{-2}$ s$^{-1}$, with $N_H$ at or above $5\times10^{22}$ cm$^{-2}$, by XMM, \Chandra, and \Suzaku.  A likely 2MASS counterpart with $K_S=10.2$ shows colors indicative of a late-type star. 
\CXOU\ is a likely counterpart to the fast hard transient \IGR.  \Chandra\ observations find $F_X$(2-10 keV)$\sim10^{-12}$ ergs cm$^{-2}$ s$^{-1}$, with large $N_H$ variations (from $2\times10^{22}$ to $>2\times10^{23}$ cm$^{-2}$).  No 2MASS counterpart is visible, to $K_S>13$.  \XMMU\ seems likely to be a new symbiotic star or symbiotic X-ray binary, while \CXOU\ is more mysterious, likely an unusual low-mass X-ray binary.

\end{abstract}

\keywords{binaries : X-rays --- stars: neutron}

\maketitle
%----------------------------------

\section{Introduction}\label{s:intro}

The wide field of view, sensitivity to hard X-rays, and program focused on the Galactic Plane of the INTEGRAL IBIS telescope has unveiled new kinds of hard X-ray transient behavior \citep{Sguera05}.  Many of these systems are high-mass X-ray binaries containing neutron stars (NSs), either eccentric Be systems where the NS feeds from an excretion disk, or supergiant fast X-ray transient systems \citep{Negueruela06} where the NS feeds from a (clumpy) stellar wind \citep{Walter07}.  A number of new INTEGRAL sources have now been identified with supergiant fast X-ray transients, showing rapid flaring behavior and variable $N_H$, and in some cases (slow) pulsations \citep{Patel04,Walter06,Chaty08}.

However, some rapidly variable X-ray transient behavior may be produced by other systems, such as symbiotic stars \citep[with white dwarf accretors, e.g.][]{Smith08}, symbiotic X-ray binaries \citep[with neutron star accretors, e.g.][]{Masetti07}, some unusual black hole X-ray binaries \citep{iZ_fxt00}, or magnetars \citep{CastroTirado08,Stefanescu08}.  Identifying the optical/infrared counterparts of rapidly variable X-ray transients is important for understanding these systems, and requires accurate positions from focusing X-ray instruments.  

The INTEGRAL hard X-ray source \IGR\ was first identified by \citet{Kretschmar04} as a hard X-ray flaring source, positionally consistent with the faint ROSAT source 2RXP J174040.9-280852.  \citet{Sguera06} described the X-ray activity as three fast (few minute timescale) bright (up to 0.8 Crab) flares, and refined the position to a 1.7' radius.  A 23 keV bremsstrahlung model was an acceptable fit to the 20-60 keV spectrum.  \citet{Sguera06} noted that \IGR's outburst was unusually strong and brief compared to known supergiant fast X-ray transients, and refrained from ruling out alternative interpretations.  

We have identified two spectrally hard, rapidly variable, X-ray transients in archival \Chandra, XMM-Newton, and \Suzaku\ data near the Galactic Center.  Our likely counterpart to the INTEGRAL IBIS fast transient \IGR\ detected in 2004, \CXOU, showed strong flaring activity in early 2007.  X-ray transient behavior has not been previously identified from the other, \XMMU, for which we identify hard X-ray flaring in late 2006 and early 2007.  

\section{XMMU J174445.5-295044}\label{s:xmm}

\subsection{Data}
We observed the supernova remnant G359.1-0.5 on Sept. 24, 2006 (ObsId 0400340101) for 42 ksec with XMM's EPIC camera, using two MOS CCD detectors \citep{Turner01} with medium filters and one pn CCD detector
\citep{Struder01} with a thin filter.  (The time resolution was 2.6 s for the MOS cameras and 73.4 ms for the pn camera.)  All data were reduced using HEASOFT 6.5\footnote{http://heasarc.gsfc.nasa.gov/docs/software/lheasoft} and SAS version 8.0.0\footnote{http://xmm2.esac.esa.int/sas/}.  The data were moderately affected by soft proton flares.  For spectral analysis, we chose a compromise between retaining photon statistics and excluding background, using limiting 0.2-10 keV rates of 5, 6, and 40 counts/s for MOS1, MOS2, and pn cameras (respectively). This left 25.0, 26.2 and 23.4 kiloseconds of time for the MOS1, MOS2, and pn cameras.  We excluded event grades higher than 12 for MOS, and higher than 4 for the pn.  The brightest point source was so much brighter than the background that all data could be used to produce lightcurves (see below).

We used \Chandra\ archival ObsIDs 658, 2278, 2289 (all with the ACIS-I imaging CCD array), and 2714 (using the HRC-I high resolution camera).  The data were reduced with CIAO 4.1 and CALDB 4.1, following standard data analysis threads\footnote{http://cxc.harvard.edu/ciao/threads}.    
We used archival XIS data from Suzaku ObsID 501056010, reduced following the Suzaku Data Reduction Guide \footnote{http://heasarc.gsfc.nasa.gov/docs/suzaku/analysis/abc/}.  We used the cleaned event files in the HEASARC archive, extracted spectra and backgrounds with XSELECT \footnote{http://heasarc.gsfc.nasa.gov/docs/software/ftools}, and produced response files with the {\it xisresp} task.  
Finally, we included archival XMM data from three previous observations of 1E 1740.7-2942, in 2001, 2003, and 2005.  These observations were conducted with the pn camera in timing mode, so only the MOS detectors were useful for our imaging.  We performed standard filtering, as discussed above.  All of these data are described in Table 1.

\subsection{Discovery of \XMMU}

XMMU J174445.5-295044 is the brightest point source in the 2006 XMM data.  We extracted its spectrum (from the data with high background excluded) and lightcurve (from the full dataset) from 20\arcsec\ circles, with background regions nearby on the same chip.  

\begin{figure}
\figurenum{1}
%\epsscale{0.6}
\includegraphics[scale=.37]{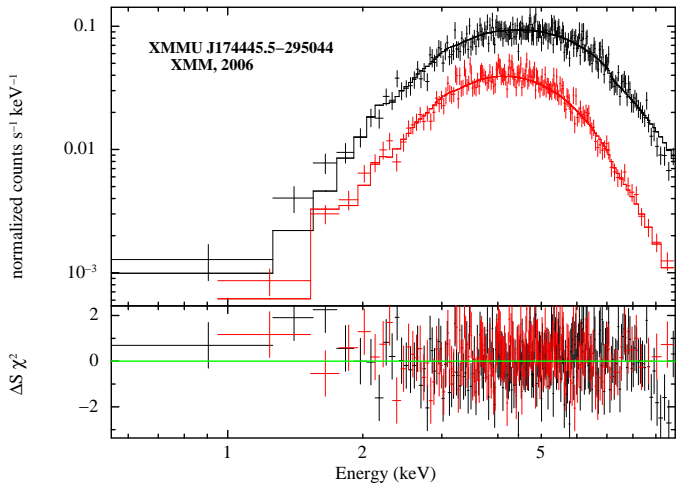}
\caption[ew1_powspec.eps]{ \label{fig:ew1_spec}
Upper panel: Spectra of \XMMU\ from 2006, with XMM EPIC pn (top) and MOS (below; red online), fit with an absorbed powerlaw (Table 2). Lower panel: residuals to the fit.
} 
\end{figure}

\XMMU\ shows a very hard absorbed spectrum (Fig. \ref{fig:ew1_spec}, Table 2), indicating it is not a nearby coronal source. A power-law fit to the pn and combined MOS data is acceptable, and finds an intrinsically hard spectrum ($\Gamma=1.18^{+0.08}_{-0.07}$) with substantial extinction ($N_H=8.6\pm0.4\times10^{22}$ cm$^{-2}$).  The spectrum is apparently featureless, with limits on the equivalent width of Fe lines at 6.4 and 6.7 keV at 26 and 9 eV respectively.  Its average unabsorbed 2-10 keV X-ray luminosity would be $1.7\times10^{35}$ ergs s$^{-1}$ if it were located at an 8 kpc distance.  

\begin{figure}
\figurenum{2}
%\epsscale{0.6}
\includegraphics[angle=0,scale=.45]{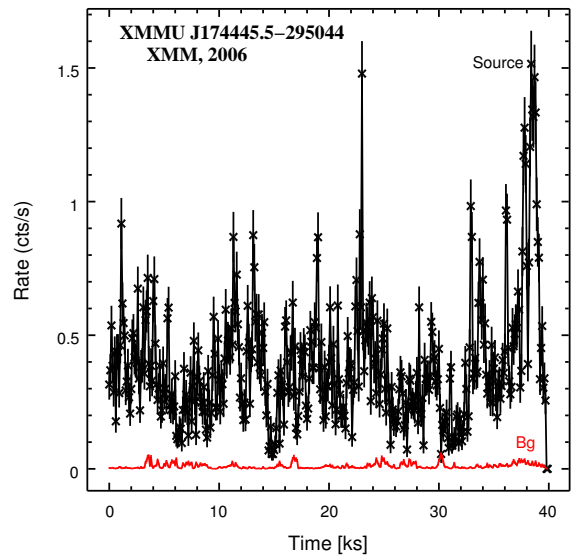}
\caption[ew1_100s_untfilt2_8bary.ps]{ \label{fig:ew1_xmm_lc}
2006 XMM pn background-subtracted 1.5-9 keV lightcurve of \XMMU, binned at 100s resolution.  Below it (in red online), we plot the background (appropriately scaled) extracted from a nearby region on the same chip.  
} 
\end{figure}

We produced barycentered lightcurves for \XMMU, and found strong flaring on timescales of hundreds of seconds (Fig. \ref{fig:ew1_xmm_lc}).  The hardness of the spectrum does not vary strongly, indicating that the flaring is intrinsic and not due to varying $N_H$. 
Power spectra, with a binning timescale of 0.4 seconds, did not produce clear evidence of any periodicity, only red noise (rms variability of 24\%, from $2\times10^{-4}$ to 0.1 Hz). 

\begin{figure}
\figurenum{3}
\includegraphics[angle=0,scale=0.45]{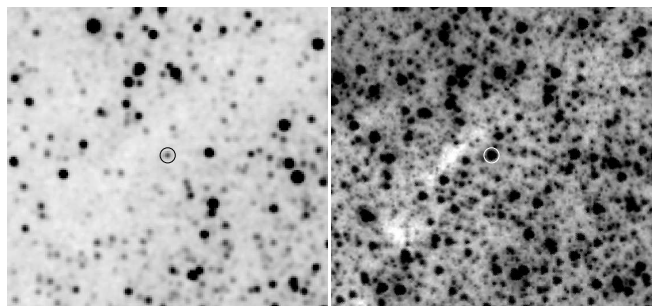}
\caption[K_GLM_XMMU.ps]{
Images in 2MASS K band (left) and GLIMPSE 4.5 micron band (right) of the \XMMU\ field.  Generous 4'' error circles are plotted around the XMM position, containing only one possible counterpart.  
\label{fig:K_XMM}
}
\end{figure}

We use an error radius of 2''\footnote{XMM-SOC-CAL-TN-0018, from http://xmm.esac.esa.int} for comparing the XMM position with accurate optical/IR catalogs (though we note that others, e.g. \citet{Walter06}, tend to use 4'').  
A potential 2MASS\footnote{see http://irsa.ipac.caltech.edu/} counterpart (2MASS J17444541-2950446) is only 1.78\arcsec\ from the XMM position, with $K_S$=10.2 and $J-K_S$=4.7 (Fig. \ref{fig:K_XMM}, Tables 3, 4).  
Only 12 2MASS stars with magnitude $K_S<10.2$ lie within 50\arcsec\ of \XMMU, indicating that the probability of finding a star of this brightness within 2\arcsec (our search radius) is only 2\%. Thus, the 2MASS star is the likely counterpart of \XMMU.

Since intrinsic $J-K$ colors range from -0.2 for O stars to 1.2 for M giants, E(J-K) must be within 3.5 to 4.9.  For standard interstellar reddening laws, $A_V=$20 to 28, and $N_H=3.6\times10^{22}$ to $5\times10^{22}$ cm$^{-2}$.
  These values suggest that if this star is the true counterpart, the X-rays probably suffer additional local absorption.  

We use the system of \citet{Neg07} to compare the near-IR colors of the 2MASS counterpart with those expected for high-mass X-ray binaries.  Following \citet{Comeron05}, we define the reddening-free parameter $Q=(J-H)-1.70(H-K_S)$.  Supergiants should have $K_S<11$ and $Q<0.2$, Be stars should have $K_S<12$ and $Q<0$, and typical late-type stars have $Q\sim0.5$.  We find $Q=0.81$ for the 2MASS counterpart, which is redder than expected even for late-type stars.  

% Late-type stars seem not to get redder than Q~0.5.
%
%If the 2MASS star is a typical Be B0V star with reddening $A_V$=28, then its distance should be only 0.9 kpc \citep{Cox00}. A supergiant B0I star would be 4.4 kpc away.  If the 2MASS star is the counterpart, then this source is closer than the Galactic Center.

\subsection{\XMMU\ at lower fluxes}

\begin{figure}
\figurenum{4}
%\epsscale{0.5}
\includegraphics[angle=0,scale=.5]{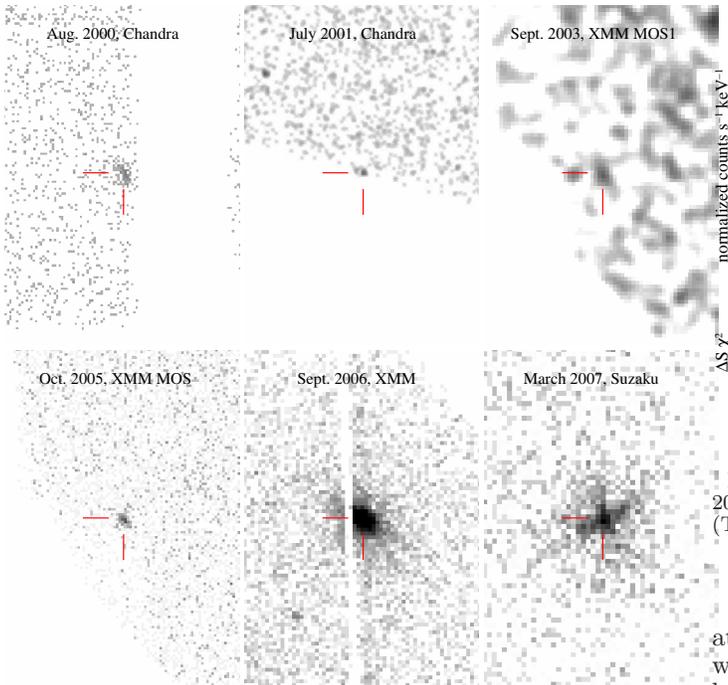}
\caption[six_XMMU_img.ps]{ \label{fig:ew1_imgs}
1.5-6 keV images of all 6 observations with detections or marginal detections of \XMMU.  All images are on the same scale (dimensions 6.3' by 8.7'), though the binning and stretch vary.  Two observations (July 2001 and Sept. 2003) have been smoothed to more clearly identify \XMMU.  The Sept. 2003 MOS observation is only at best a marginal detection.  Note that four of the observations place \XMMU\ near or at the edge of the detector.  
} 
\end{figure}

The location of \XMMU\ ($l$=359.13, $b$=-0.31), at the edge of the \Chandra\ wide Galactic Center survey field \citep{Wang02}, and within 15' of the LMXB 1E 1740.7-2942, has been observed numerous times by X-ray satellites.  We have searched the HEASARC archive\footnote{http://heasarc.gsfc.nasa.gov/db-perl/W3Browse/w3browse.pl} for detections or constraining upper limits.  We report here archival detections of \XMMU\ at lower fluxes by Suzaku, XMM, and \Chandra, and upper limits from \Chandra\ and XMM (Table 2, Fig. \ref{fig:ew1_imgs}).  Observations by ROSAT, ASCA, EXOSAT, and Einstein do not detect it, but the upper limits are not particularly constraining, so we do not discuss those observations further.

\XMMU\ is consistent with the position of CXOGC J174445.5-295042 \citep{Muno06}, seen in \Chandra\ ObsID 2278. (The Chandra position is at the edge of a chip, so its positional uncertainty of 2.8'', while consistent with \XMMU, may not be fully trustworthy.)  We have attempted spectral fitting to an absorbed power-law and absorbed bremsstrahlung spectrum, using the unbinned data and the C statistic (Table 2).  Testing the acceptability of a C-statistic fit is done through Monte Carlo simulations (we use 1000) of the model.  The percentage of those simulations having a larger C-statistic than the data is reported as the 'goodness'; a high value (e.g. 95\%) indicates a poor fit.  Its 10-count detection at the very edge of the chip makes conclusions tentative.  

\begin{figure}
\figurenum{5}
%\epsscale{0.6}
\includegraphics[scale=.35]{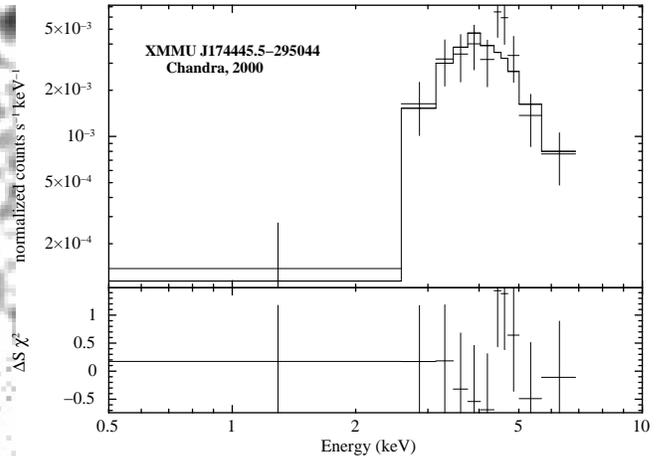}
\caption[ew1_658_spec.eps]{ \label{fig:658_spec}
\Chandra\ spectrum of \XMMU, from 2000 data (ObsID 658), fitted with an absorbed powerlaw model (Table 2).
} 
\end{figure}

    A second \Chandra\ ACIS-I observation (ObsID 658) identifies \XMMU\ on the off-axis S2 chip, with over 100 counts.  We extracted a spectrum (binning the data with 10 counts/bin) and a lightcurve with 200s bins.  Chi-square and KS tests found no evidence of variability in the lightcurve.  We fit the spectrum (Fig. \ref{fig:658_spec}) with absorbed power-law and bremsstrahlung models (Table 2). Fitting the unbinned data with the C-statistic gave similar results (within the errors). 

A third \Chandra\ ACIS-I observation (ObsID 2289) did not detect \XMMU, at the extreme corner of the off-axis S2 chip.  Neither did a \Chandra\ HRC-I observation where \XMMU\ was far off-axis (ObsID 2714).  To determine upper limits for nondetections of \XMMU\  in this paper, we assume a spectrum similar to its faintest detections; an absorbed power-law with photon index 2 and $N_H=10^{23}$ cm$^{-2}$.  We extract the number of photons in a circle corresponding to the 50\% encircled energy radius for photons of 4.5 keV at \XMMU's off-axis angle, using an energy range of 1.5-6 keV.  Using a larger nearby background region free of sources, we compute the expected number of background photons in this circle, scale it to the source region, and compute the 90\% confidence lower limit \citep{Gehrels86} on the number of expected background photons in the source region.  The upper limit on the source countrate is derived by subtracting this lower limit from the observed photons, accounting for the encircled energy fraction and the vignetting of the telescope.  Finally, we calculate flux upper limits using PIMMS\footnote{http://asc.harvard.edu/toolkit/pimms.jsp}.  

\begin{figure}
\figurenum{6}
%\epsscale{0.6}
\includegraphics[scale=.34]{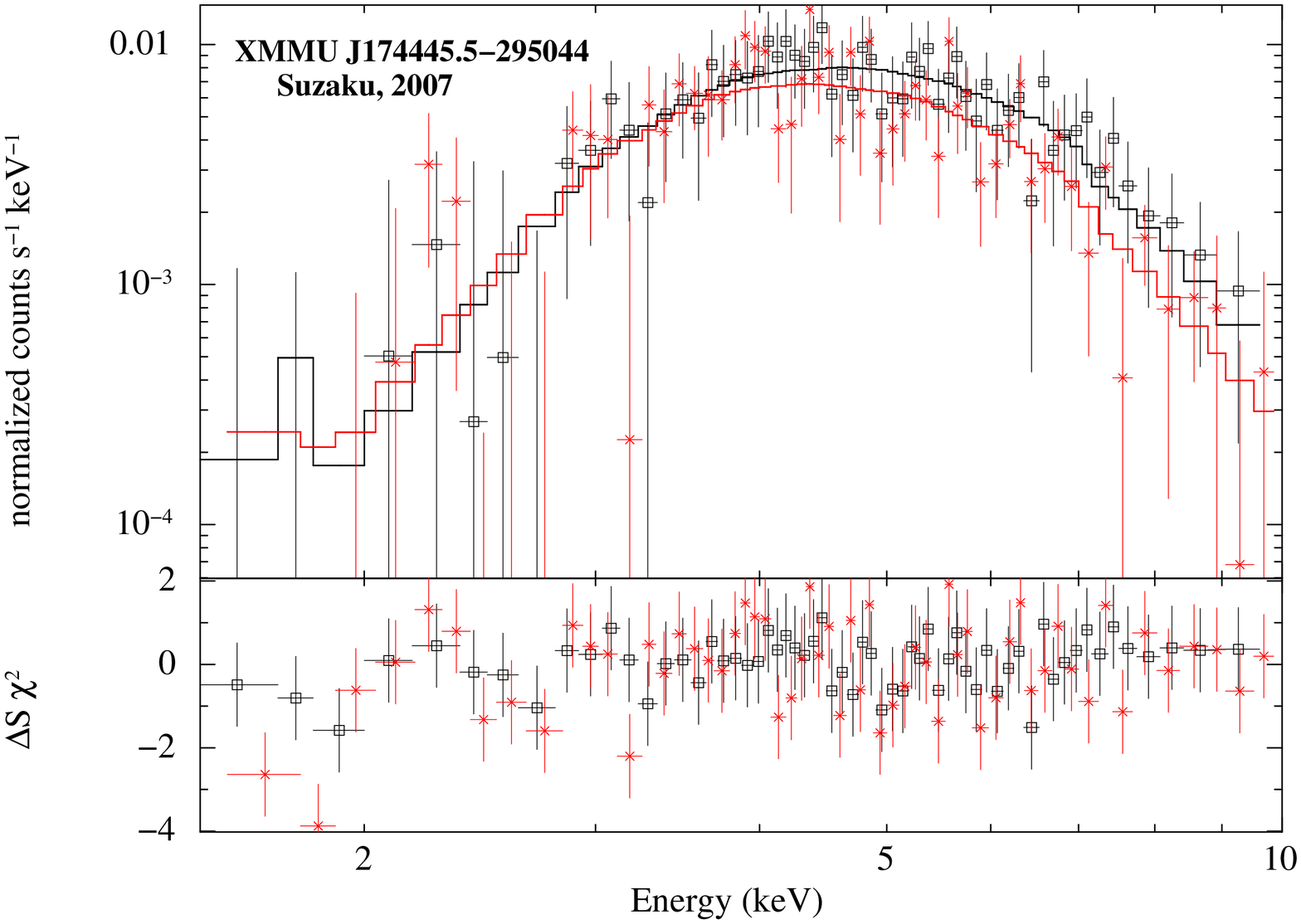}
\caption[ew1_suzxis_spec.ps]{ \label{fig:ew1_suz_spec}
\Suzaku\ spectrum of \XMMU, fitted with an absorbed powerlaw model (fit only above 2 keV).  XIS1 data (diagonal crosses, red online) are lower above $4$ keV than the combined XIS0 and XIS3 data (boxes, black online).
} 
\end{figure}

\begin{figure}
\figurenum{7}
%\epsscale{0.6}
\includegraphics[angle=0,scale=.45]{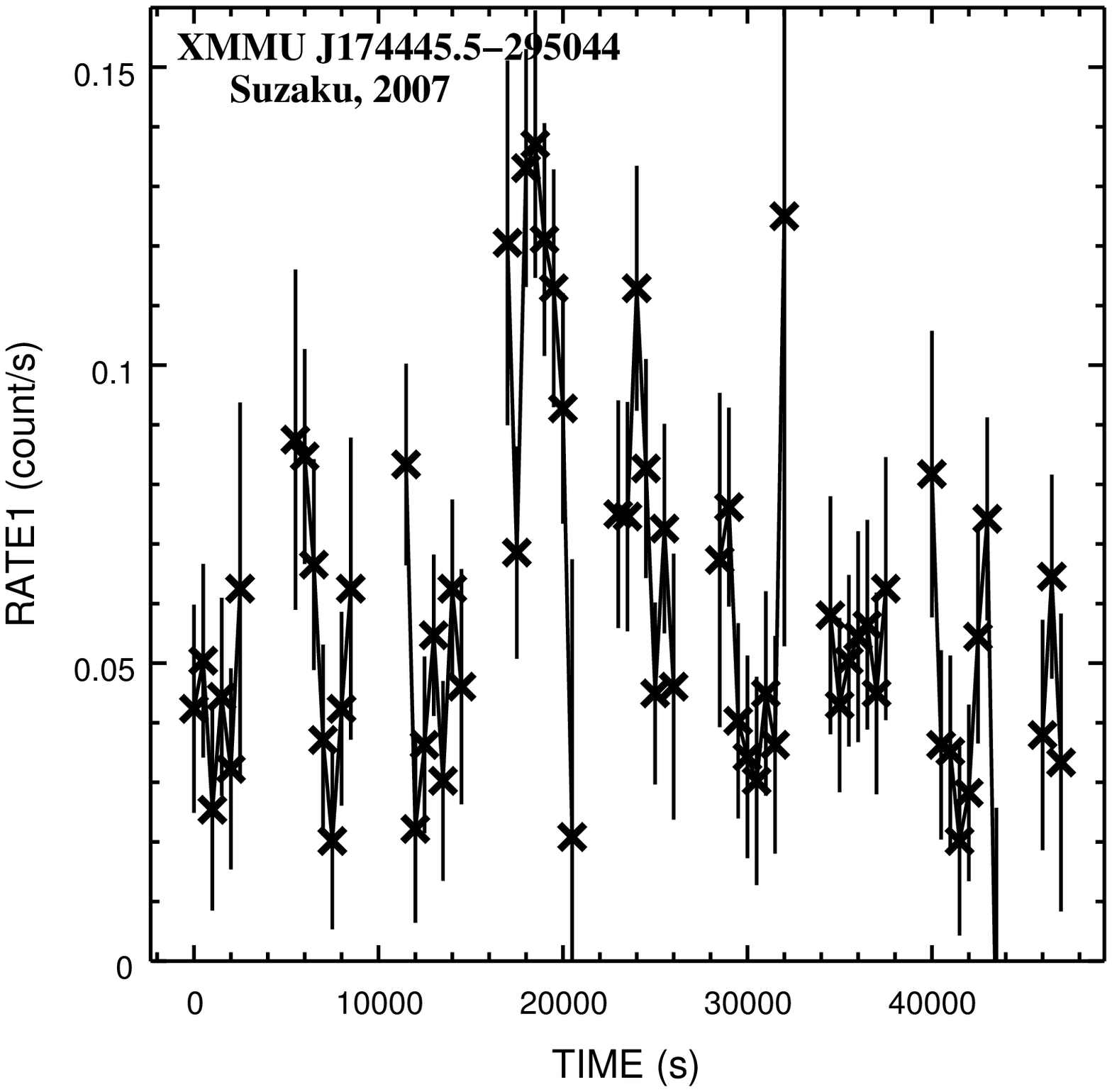}
\caption[xis_500s_sub.eps]{ \label{fig:ew1_suz_lc}
\Suzaku\ combined XIS 1.5-6 keV lightcurve of \XMMU, using 500s bins.
} 
\end{figure}

%\clearpage

A recently archived \Suzaku\ observation (ObsID 501056010, 2007) clearly detects \XMMU\ in all three operational XIS \citep{Koyama_XIS07} cameras (due to source confusion in the Galactic Center, we do not consider the PIN data here).  We extract spectra from a 100\arcsec\ region of the cleaned data, combining the spectra from XIS CCDs 0 and 3, and binning the data at 40 photons/bin (Fig. \ref{fig:ew1_suz_spec}).  The variations in thermal emission from the Galactic Center and large PSF of \Suzaku\ make background subtraction difficult below 2 keV.  Since \XMMU\ shows little emission below 2 keV, we choose to ignore data below 2 keV for spectral fitting.  The flux is a factor of 5 lower than observed by XMM, with a softer spectrum, and slightly higher $N_H$.  We created a lightcurve from all detectors, in the energy range 1.5-6 keV, at 16 s binning.  Variability, suggesting continued flaring, is best seen at 500 s binning (Fig. \ref{fig:ew1_suz_lc}).

We retrieved three prior XMM observations of 1E1740.7-2942 from the archive (Table 1), taken in 2001, 2003, and 2005.  In each case, the pn camera was in timing or small window mode, so we used only MOS data.  The 2001 and 2003 data were heavily affected by background flaring, forcing us to use partly contaminated data.  A combined MOS image of the 2001 data produced only an upper limit on the existence of \XMMU.  The 2005 data produced a clear detection of \XMMU\ (Fig. \ref{fig:ew1_imgs}, lower left), with a flux comparable to the \Chandra\ detections (Table 2).   In the 2003 data, the position of \XMMU\ falls off the MOS2 chips, but lies on a MOS1 chip.  Although the source is not detected with a blind detection algorithm, our upper limit method described above calculates $11^{+2.2}_{-4.8}$ counts (at 90\% confidence).  Thus we consider this as a marginal detection (see Fig. \ref{fig:ew1_imgs}, upper right).  

\section{\CXOU: An INTEGRAL Source?}\label{s:igr}

\begin{figure}
\figurenum{8}
%\epsscale{0.6}
\includegraphics[angle=0,scale=.47]{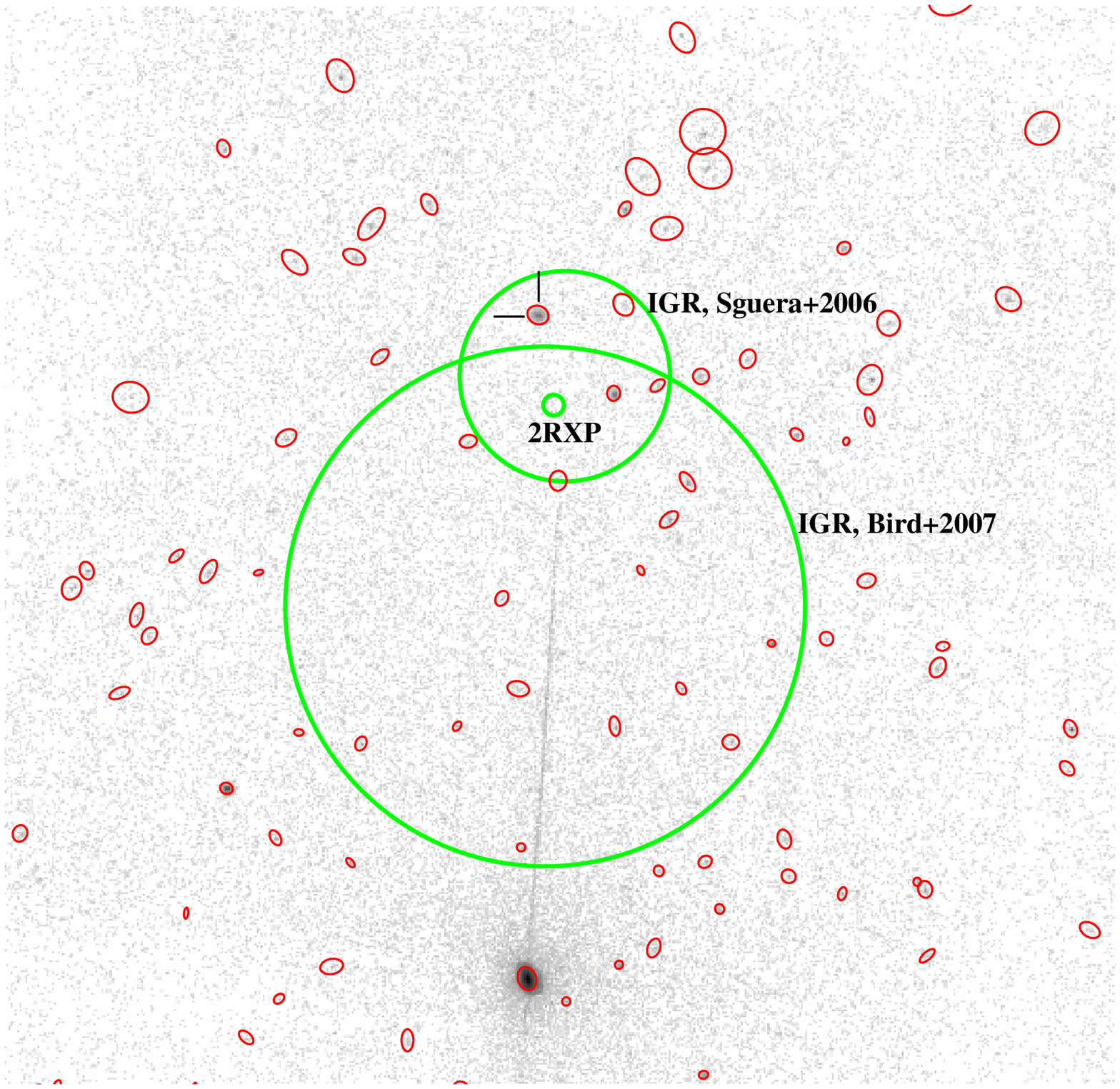}
\caption[mosaic.ps]{ \label{fig:igr_mosaic}
Mosaic of five \Chandra\ observations in the 0.5-7 keV range, showing the large (green online) error circles for \IGR from \citet{Bird07} and \citet{Sguera06}, and (red online) ellipses around detected sources.  The candidate counterpart \CXOU\ is indicated by two black ticks.  The small thick (green online) error circle indicates the location of 2RXP J174040.9-280852; note that no \Chandra\ sources lie near it.  SLX 1737-282 is the bright source producing a readout streak, near the bottom of the image.
} \end{figure}

%\subsection{\Chandra\ Data Analysis}
\citet{Sguera06} give the revised INTEGRAL error circle as R.A.=$17^h40^m40^s.08$, decl.=$-28^d 08' 24''$ (J2000), with error radius of 1.7'.  \citet{Bird07} gives the error circle as $17^h40^m42^s$, $-28^d12.1'$ , with a 4.2' error radius.  
The Galactic Bulge Latitude Survey (Grindlay et al., in prep.) covers both error circles thoroughly, between 4 overlapping 15 ks ACIS-I observations.  We downloaded these four observations, plus the ACIS-S observation of \citet{Tomsick08}.  We searched for periods of enhanced background, but found none.  We created a mosaic image in the 0.5-7 keV energy band (Fig. \ref{fig:igr_mosaic}). 

The mosaic image shows eighteen sources in one or the other of the two overlapping error circles.  (Note that no \Chandra\ source is visible in the error circle for the ROSAT source 2RXP J174040.9-280852.)  The brightest of these sources, CXOU J174042.0-280724 (indicated), shows marked flaring variability and a hard spectrum showing strong variations in absorption, which suggest that this source is the soft X-ray counterpart to \IGR.  None of the other sources within the \IGR\ error circles show these behaviors, so we focus on CXOU J174042.0-280724. (Tomsick et al. (2008) identify the brightest source visible in Fig. \ref{fig:igr_mosaic} as the low-mass X-ray binary SLX 1737-282, which is not a likely counterpart for \IGR.)

CXOU J174042.0-280724 is in the field of view of ObsIDs 8199, 8200 (on the edge of the chip), 8202, and 7526, all of which were observed between July 30 and August 1, 2007 (Table 1).  The count rate is much higher in ObsID 8202 than the others, by almost a factor of 10.  None of these datasets showed background flaring, so we used all the data, and extracted spectra and lightcurves using CIAO tools. 
% 8202: 370/14553=0.025
% 8199: 58/14682= 0.0039
% 8200: 24/14556= 0.0016  (but partly off chip)
% 7526: 15/5108=  0.0029

\begin{figure}
\figurenum{9}
%\epsscale{0.5}
\includegraphics[scale=.38]{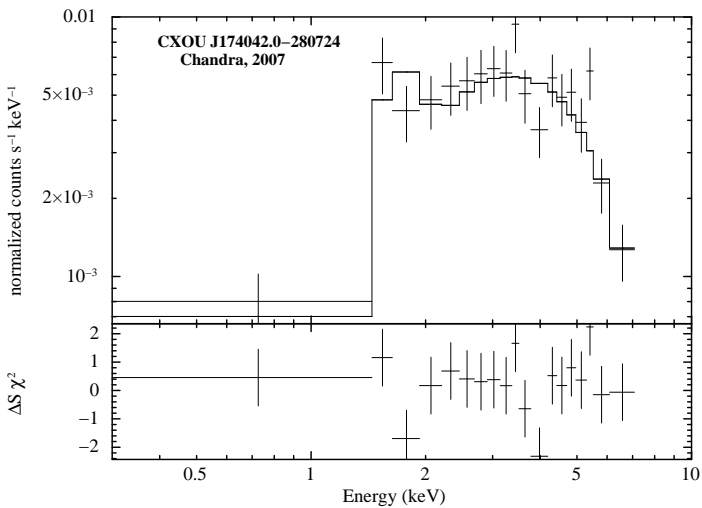}
\caption[8202a_spec.ps]{ \label{fig:8202_spec}
Spectrum of \CXOU\ in \Chandra\ ObsID 8202, fit with an absorbed bremsstrahlung spectrum.
} 
\end{figure}

\begin{figure}
\figurenum{10}
%\epsscale{0.6}
\includegraphics[angle=0,scale=.43]{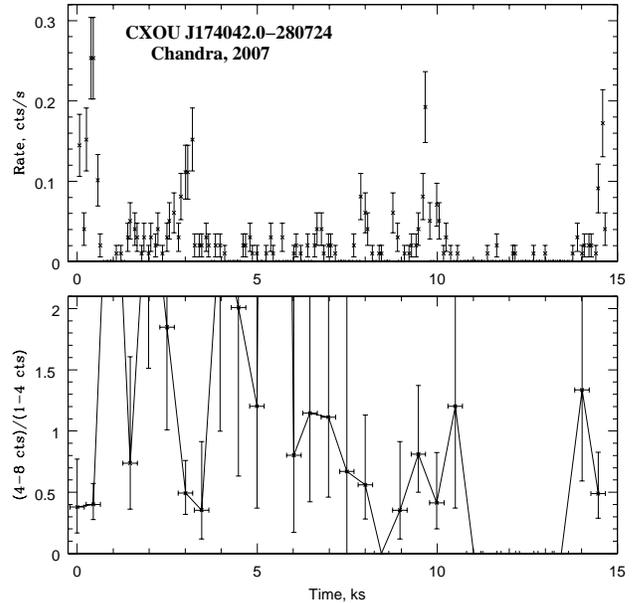}
\caption[8202a_100s.eps]{ \label{fig:8202_lc}
Top: Lightcurve of \CXOU\ in \Chandra\ ObsID 8202, 0.5-7 keV, binned at 100 s resolution.  Bottom: Hardness ratio of \CXOU\ in ObsID 8202, 4-8 keV counts over 1-4 keV counts, in 500 s bins.  
} 
\end{figure}

A power-law spectral fit of CXOU J174042.0-280724 in ObsID 8202 finds a hard, absorbed spectrum, with photon index $\Gamma=0.9^{+0.4}_{-0.6}$ and $N_H=1.7^{+0.7}_{-0.9}\times10^{22}$ cm$^{-2}$ (Table 5; Fig. \ref{fig:8202_spec}).  A lightcurve (binned at 100 s) shows irregular sharp, short flares (Fig. \ref{fig:8202_lc}a), reaching at least 100 times the lowest flux.   No periodicities are detected in the overall power spectrum.  Although the count rates between flares are low, there is some indication that part of the variability is due to variable absorption, as the emission is systematically softer during the brightest periods (Fig. \ref{fig:8202_lc}).

We test this by extracting spectra from bright and faint parts of ObsID 8202, and fitting the (ungrouped) spectra with an absorbed power-law model, with photon index constrained between 0 and 3. 
Our spectral fit to the brightest 220 seconds of the first flare finds $N_H=0.8^{+1.5}_{-0.7}\times10^{22}$ cm$^{-2}$, and a 2-10 keV flux of $F_X=8^{+4}_{-2}\times10^{-12}$ ergs cm$^{-2}$  s$^{-1}$.  (We note that there is no evidence for pileup in the spectrum or image of this flare, due to the source's off-axis location). This is still three orders of magnitude fainter than the peak INTEGRAL flux from \IGR.  A similar fit to the data below 0.04 counts/s (excluding the last 5 ksec) gives $N_H=2.0^{+2.4}_{-1.6}\times10^{22}$ cm$^{-2}$, and a flux of $F_X=3.6^{+2.2}_{-1.3}\times10^{-13}$ ergs cm$^{-2}$ s$^{-1}$.  This exercise strongly supports intrinsic source variation, and does not prove $N_H$ variations during ObsID 8202.

\begin{figure}
\figurenum{11}
%\epsscale{0.6}
\includegraphics[scale=.35]{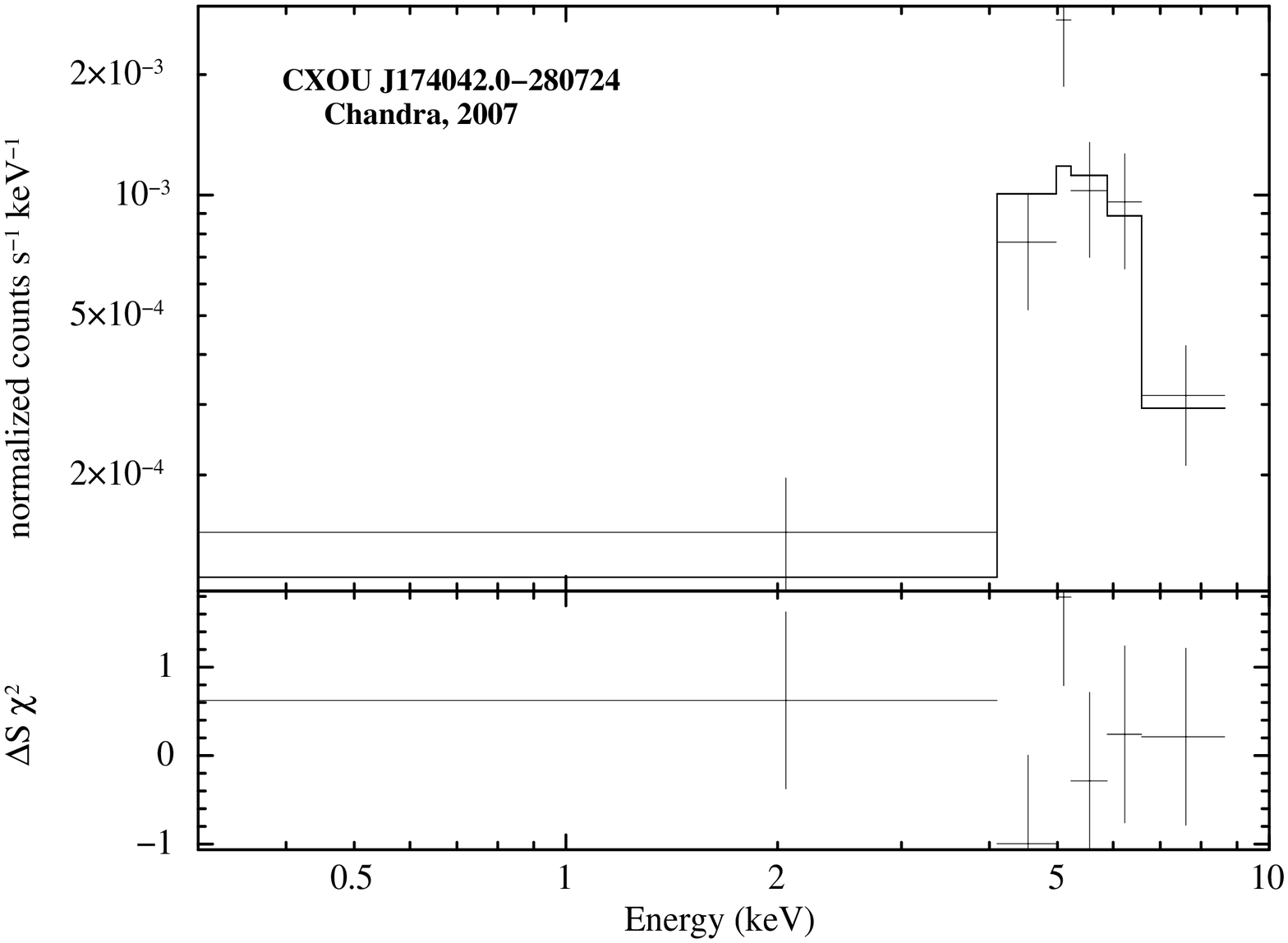}
\caption[8199a_spec.ps]{ \label{fig:8199_spec}
Spectrum of \CXOU\ in \Chandra\ ObsID 8199, fit with an absorbed bremsstrahlung spectrum.  Note the higher low-energy cutoff in comparison with Fig. \ref{fig:8202_spec}. 
} 
\end{figure}

\begin{figure}
\figurenum{12}
%\epsscale{0.6}
\includegraphics[angle=0,scale=.45]{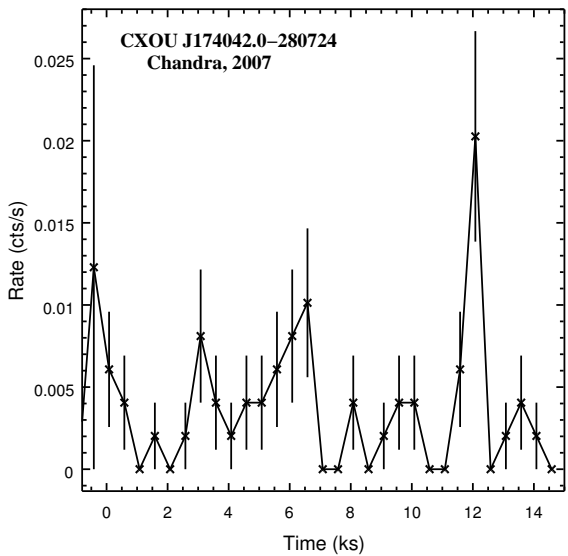}
\caption[8199a_100s.eps]{ \label{fig:8199_lc}
Lightcurve of \CXOU\ in \Chandra\ ObsID 8199, 0.5-7 keV, binned at 500s resolution.
} 
\end{figure}

\CXOU's substantially smaller count rate in the other three observations is explained by comparison of the spectra from ObsIDs 8199 and 8202.  The spectrum observed in 8199 is substantially harder than that in 8202, showing a low-energy cutoff near 4 keV vs. 1.5 keV (compare Fig. \ref{fig:8202_spec} vs. \ref{fig:8199_spec}).  Fitting bremsstrahlung models to both spectra confirms that the $N_H$ value changed dramatically between the two observations, from $N_H=29^{+36}_{-7}\times10^{22}$ cm$^{-2}$ to $N_H=2.2^{+0.5}_{-0.3}\times10^{22}$ cm$^{-2}$, while the unabsorbed 2-8 keV fluxes are consistent (Table 5).  Thus the inter-observation variability may be ascribed entirely to varying $N_H$, although the flaring (present in ObsID 8199 also; Fig. \ref{fig:8199_lc}) cannot.   The remaining two observations (ObsIDs 8200 and 7526) offer less information (in 8200 the source lies on a chip edge, while 7526 was a shorter observation), producing spectra which are poorly constrained.  However, the spectrum from ObsID 8200 is as hard as that from 8199 (only 3 of 27 photons are below 4 keV in ObsID 8200, compared to 7 of 63 in ObsID 8199), giving support to our explanation of the variability.  The spectrum from ObsID 7526 is rather softer (7 of 19 counts below 4 keV), which might be due to intrinsic spectral evolution, or to $N_H$ variations.  Future observations could probe \CXOU's spectral evolution.  

%(Identify how many counts???) PI between 0, 274; 274,685.
% 8200: 3 cts below 4 keV, 24 in 4-10 keV.
% 8199: 7 below, 56 above.
% 8202: 233 below, 173 above.
% 7526: 8 below, 14 above. -> 6.9 below,12.5 above.
% 7526_bg: 120,162:1.1,1.5.

%\begin{figure}
%\figurenum{13}
%\includegraphics[angle=0,scale=0.42]{K_GLM_images.ps}
%\caption[K_GLM_images.ps]{
%Images of the field of \CXOU\ from the 2MASS K band (left) and GLIMPSE 3.6 micron band (right).  A generous 1'' \Chandra\ error circle is plotted, enclosing no sources.
%\label{fig:K_CXOU}
%}
%\end{figure}

The best Chandra position for \CXOU\ is RA=17:40:42.05$\pm0.01$,  Dec=-28:07:24.6$\pm0.1$.  Without astrometric matching of nearby X-ray sources to other wavelengths, we must add a systematic uncertainty of 0.6'' (90\% conf.).  We have searched for possible counterparts in the 2MASS catalog, but the nearest star is 1.9'' away, at a magnitude of 11.8.
%(Fig. \ref{fig:K_CXOU})
  Based on the faintest 2MASS sources detected within 1', rough limits of $J>14$ and $K_S>13$ can be set for the error circle.  No counterparts exist in SIMBAD or the USNO catalog.

\section{Discussion}

\begin{figure}
\figurenum{13}
\includegraphics[angle=0,scale=.43]{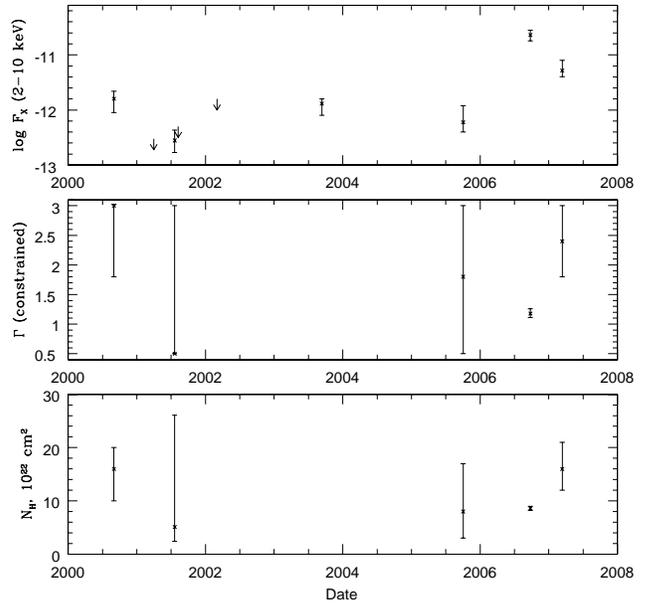}
\caption{\label{fig:longterm}
Fluxes, photon indices, and $N_H$ measurements for \XMMU\ from constrained ($0.5 < \Gamma < 3.0$) absorbed power-law spectral fits to XMM, \Chandra\ and \Suzaku\ data over 8 years (see Table 2).  }
\end{figure}

We show the fluxes, photon indices, and $N_H$ measurements for \XMMU\ in Fig. \ref{fig:longterm}.  Clearly \XMMU\ spends most of its time at lower fluxes, below $F_X=3\times 10^{-12}$ ergs cm$^{2}$ s$^{-1}$.  The 2006 XMM and 2007 \Suzaku\ observations clearly show differences in photon index and $N_H$.  The \Chandra\ results also suggest a softening of the intrinsic spectrum, and increased $N_H$, at lower fluxes.   The variable absorption and hard spectrum seem to rule out a coronal nature for the X-ray emission, suggesting an accreting compact object. 

The evidence of rapid variability on timescales of minutes, the hard spectrum,  and the intrinsic and variable absorption suggest a high-mass X-ray binary nature for \XMMU\ \citep{White95}. 
However, the red colors of the infrared counterpart for \XMMU\ are inconsistent with a high-mass X-ray binary nature, suggesting a red giant companion.  
The spectrum and variability are similar to those of, e.g., 4U 1954+31 \citep{Masetti07}, suggesting a symbiotic X-ray binary nature.  However, some symbiotic stars also show hard spectra and rapid variability \citep{Luna07}, so a white dwarf accretor is also a viable explanation. 
We note that there are strong similarities between accretion from the clumpy stellar wind of high-mass stars vs. that from giant stars onto a compact object.  
Indeed, several INTEGRAL sources with X-ray properties similar to high-mass X-ray binaries have recently been identified as symbiotic X-ray binaries \citep{Masetti07,Nespoli08,Tomsick08}.

The substantial extinction to \XMMU's infrared counterpart suggests a distance of at least a kpc, and thus X-ray luminosities of $>10^{32}$ ergs s$^{-1}$ in quiescence, and $>10^{33}$ ergs s$^{-1}$ at maximum.  
The flux range and strong variability of \XMMU\ suggest it may fit into the category of very faint X-ray transients \citep{Wijnands06,Degenaar08}, although it may reach higher fluxes at other times.  Inspection of the ASM and RXTE bulge scan data finds no evidence of outbursts from this source, with the bulge scan providing typical upper limits of 2-10 mCrab (C.\ Markwardt, priv.\ comm, 2008; R.\ Remillard, priv.\ comm, 2008).  Some of the other known very faint X-ray transients may be symbiotic systems; this possibility might be ruled out with deep infrared observations.  
%For 8 kpc, this gives $L_X$(2-10)$\lesssim$0.8-$2\times10^{36}$ ergs/s .  

\CXOU\ has a more mysterious nature than \XMMU, due to its lack of an optical counterpart.  Even if it were not associated with \IGR\ (which we think it is), its varying $N_H$ and rapid variability suggest accretion onto a compact object.  
The high flux attained by \IGR\ indicates a neutron star or black hole nature.  
  The distance to this source is unknown, as we do not have a measurement of the interstellar reddening, and at least some of the observed (variable) absorption must be intrinsic.  
%% A_Ks=0.683, A_V=7.67, N_H=1.4e22
This direction ($l$=0.12, $b$=1.35) is sufficiently reddened ($N_H$ up to $1.4\times10^{22}$ cm$^{-2}$, \citet{Marshall06}) that interstellar reddening could explain the observed $N_H$ in ObsID 8202, which has the lowest observed $N_H$ value.  
The $K_S$ limit implies $M_K > -2.2$ for a distance of 10 kpc, excluding main sequence stars with spectral types earlier than B1 and giant stars with spectral types later than K2.  At a distance of 15 kpc, these limits shift to excluding main sequence stars earlier than O9 and giants later than K4, while supergiants are ruled out for either distance.  

Some low- or intermediate-mass X-ray binaries, such as V4641 Sgr \citep{iZ_fxt00,Revnivtsev02,Chaty03} or XTE J1901+014 \citep{Smith07,Karasev07}, show rapid flaring similar to that seen from \IGR.  Alternatively, this object could be a Be-type binary at a distance over 10 kpc.  We note that for a distance of 10 kpc, the peak INTEGRAL flux corresponds to $L_X$(20-60 keV)=$1.1\times10^{38}$ ergs s$^{-1}$, suggesting a bolometric flux above the Eddington limit for a neutron star.

Further studies of these mysterious sources are strongly encouraged to improve our understanding of their nature and of fast X-ray transient behavior.  The most critical need is for deep optical and infrared imaging to search for \CXOU's expected donor star.  Spectroscopy of such an object (if possible) would be very valuable.  Optical and/or infrared spectroscopy of \XMMU's optical counterpart is also needed, to identify its spectral type and surface gravity and search for evidence of accretion.  This will test the hypothesis presented here that \XMMU\ is a symbiotic system.  If the distance can be constrained, the nature of the accreting object (white dwarf or neutron star?) may be determined.   Additional X-ray observations of both objects will also be useful to monitor their long-term activity, especially if \CXOU\ produces additional outbursts detectable by INTEGRAL.

\acknowledgements

COH thanks Craig Markwardt and Ron Remillard for their rapid efforts to search the RXTE bulge scans and RXTE/ASM records, respectively.  
COH acknowledges support by NSERC and NASA XMM GOF funding.  
JAT acknowledges partial support from {\em Chandra} award number GO8-9055X issued by the {\em Chandra X-Ray Observatory Center}, which is operated by the Smithsonian Astrophysical Observatory for and on behalf of the National Aeronautics 
and Space Administration (NASA), under contract NAS8-03060.  
This publication makes use of data products from the Two Micron All Sky Survey, which is a joint project of the University of Massachusetts and IPAC/Caltech, funded by NASA and the NSF.  This research has made use of the NASA/IPAC Infrared Science Archive, and the Spitzer Space Telescope, which are operated by the Jet Propulsion Laboratory, Caltech, under contract with NASA.  This research has made use of data obtained from the High Energy Astrophysics Science Archive Research Center (HEASARC), provided by NASA's Goddard Space Flight Center.  It is based in part on observations obtained with XMM-Newton, an ESA science mission with instruments and contributions directly funded by ESA member states and NASA.

\bibliography{fast_trans}
\bibliographystyle{apj}

\begin{deluxetable}{lcccr}
\tabletypesize{\footnotesize}
\tablewidth{7truein}
\tablecaption{\textbf{X-ray observations of \XMMU\ and \CXOU}}
\tablehead{
\colhead{Mission/Instrument} & \colhead{ObsID} & \colhead{Date} &  \colhead{Exposure} & \colhead{Notes} 
}
\startdata
\multicolumn{5}{c}{\XMMU} \\
\hline
\Chandra/ACIS-I &  658 & 2000-08-30 & 9235  &  \\
XMM/MOS   & 0112971701 & 2001-03-31 & 7003  & Nondetection \\
\Chandra/ACIS-I & 2278 & 2001-07-20 & 11611 & At chip edge  \\
\Chandra/ACIS-I & 2289 & 2001-07-21 & 11611 & Nondetection \\
\Chandra/HRC-I  & 2714 & 2002-03-02 & 29709 & Nondetection \\
XMM/MOS   & 0144630101 & 2003-09-11 & 4477  & Marginal detection \\
XMM/MOS   & 0303210201 & 2005-10-02 & 21823  &   \\
XMM/MOS & 0400340101 & 2006-09-24 & 41028(25026)  & Source flaring  \\
XMM/pn  & 0400340101 & 2006-09-24 & 35337(23378)  & Source flaring  \\
Suzaku/XIS & 501056010 & 2007-03-13 & 26545 & Variable  \\
\hline
\multicolumn{5}{c}{\CXOU} \\
\hline
\Chandra/ACIS-I & 8199 & 2007-07-30, 01:31 & 14682 & Source flaring \\
\Chandra/ACIS-I & 8200 & 2007-07-30, 05:57 & 14566 & At chip edge \\
\Chandra/ACIS-I & 8202 & 2007-07-30, 14:41 & 14300 & Source flaring  \\
\Chandra/ACIS-S & 7526 & 2007-08-01,  13:04 & 5108  &  \\
\enddata
\tablecomments{For ObsID 0400340101, we give the exposure with background flares removed in parentheses (these data are used for spectral analysis).  
\label{tab:obs}}
\end{deluxetable}

\begin{deluxetable}{lcccccccr}
\tabletypesize{\footnotesize}
\tablewidth{7.4truein}
\tablecaption{\textbf{Spectral fits to \XMMU}}
\tablehead{
\colhead{Mission/Instrument} &  \colhead{ObsID} & \colhead{Flux} & \colhead{$\Gamma$} & \colhead{$N_H$} & \colhead{$\chi^2$/dof} & \colhead{kT} & \colhead{$N_H$} & \colhead{$\chi^2$/dof} \\
  &   & ergs cm$^{-2}$ s$^{-1}$ &  & $10^{22}$ cm$^{-2}$ &  & keV & $10^{22}$ cm$^{-2}$ &  
}
\startdata
 &  & \multicolumn{4}{c}{---------------------------Power-law fit------------------------------} & \multicolumn{3}{c}{----------Bremsstrahlung fit-----------} \\
\hline
\Chandra/ACIS-I  & 658 & $1.6^{+0.6}_{-0.7}\times10^{-12}$ & $3.0^{+0 a}_{-1.1}$ & $15^{+3}_{-6}$ & 0.68/10 & $1.8^{+5.0}_{-1.0}$ & $19^{+14}_{-9}$ & 0.57/10  \\ %ReDone
XMM/MOS   & 0112971701 & $<3\times10^{-13}$  \\
\Chandra/ACIS-I  & 2278 & $2.8^{+1.5}_{-1.1}\times10^{-13}$ & $0.5^{+2.5 a}_{-0 a}$ & $5.1^{+21}_{-2.7}$ & 146/15$^b$ & $>2.3$ & $11^{+8}_{-7}$ & 146/85$^b$ \\ %redone
\Chandra/ACIS-I & 2289 & $<5\times10^{-13}$ &  \\
\Chandra/HRC-I  & 2714 & $<1.6\times10^{-12}$ &  \\
XMM/MOS   & 0144630101 & $1.3^{+0.3}_{-0.5}\times10^{-12}$ & \\ %done
XMM/MOS   & 0303210201 & $6^{+6}_{-2}\times10^{-13}$ & $1.8^{+1.2 a}_{-1.3 a}$ & $8^{+9}_{-5}$ & 0.54/11 & $>1.8$ & $7.8^{+9.5}_{-3.3}$ & 0.54/11  \\ %done
XMM/EPIC  & 0400340101 & $2.29\pm0.05\times10^{-11}$ & $1.18^{+0.08}_{-0.07}$ & $8.6^{+0.4}_{-0.4}$ & 0.91/368 & $>126$ & $8.8^{+0.2}_{-0.2}$ & 0.92/368   \\ %Done 
Suzaku/XIS & 501056010 & $5.2^{+2.8}_{-1.2}\times10^{-12}$ & $2.4^{+0.6 a}_{-0.6}$ & $16^{+5}_{-4}$ & 0.65/111 & $6.6^{+11}_{-2.9}$  & $13.9^{+3.9}_{-3.2}$ & 0.66/111 \\ %done
\enddata
\tablecomments{Photon indices are restricted to the physically reasonable range of 0.5 to 3.  Fluxes are unabsorbed, for 2-10 keV, using best-fit absorbed power-law model, or a power-law spectrum with $\Gamma=2$ and $N_H=1\times10^{23}$ cm$^{-2}$ for upper limits.  \\
$^a$: Error calculation reached hard limit.\\
$^b$: Used C-statistic, reporting C-statistic and ``goodness'' (see text) instead of $\chi^2$ and degrees of freedom.
\label{tab:xmmu}}
\end{deluxetable}

\begin{deluxetable}{lcccr}
\tabletypesize{\footnotesize}
\tablewidth{4.5truein}
\tablecaption{\textbf{Positions of \XMMU, \CXOU, and possible counterparts}}
\tablehead{
\colhead{Name} & \colhead{RA} & \colhead{Dec} & \colhead{$\Delta$}  & \colhead{Notes} 
}
\startdata
\XMMU\ & 17:44:45.55 & -29:50:44.3 & $\pm2.0^a$  & XMM pos. \\
\XMMU\ & 17:44:45.54 & -29:50:42.1 & $\pm2.8^b$    &   Muno+06 \\
2MASS J17444541-2950446 & 17:44:45.41 & 29:50:44.6 & $\pm0.07^a$ &   \\
\hline \\
\CXOU\ & 17:40:42.05 & -28:07:24.6 & $\pm0.6^b$ & CXO pos. \\
\enddata
\tablecomments{Positions of detected (X-ray or optical/IR) source and error.  $^a$: 1$\sigma$ error.  $^b$: 90\% confidence error.
\label{tab:pos}}
\end{deluxetable}

\begin{deluxetable}{lcr}
\tabletypesize{\footnotesize}
\tablewidth{3.5truein}
\tablecaption{\textbf{Magnitudes and Fluxes of 2MASS J17444541-2950446}}
\tablehead{
\colhead{Band} & \colhead{Magnitude} & \colhead{Flux}  
}
\startdata
$J$   & 14.89(5) & 1.77   \\
$H$   & 11.62(4) & 23.1   \\
$K_S$ & 10.17(3) & 57.2   \\
3.6$\mu$m & 9.06(5) & 66(3)  \\
4.5$\mu$m & 8.87(5) & 51(2)  \\
5.8$\mu$m & 8.67(4) & 40(2)  \\
8.0$\mu$m & 8.67(5) & 22(1)  \\
\enddata
\tablecomments{Fluxes in mJy.  Magnitudes from 2MASS and GLIMPSE.
\label{tab:ir}}
\end{deluxetable}
%  17444541-2950446 266.189228 -29.845734    17h44m45.41s   29d50m44.64s    0.07    0.06       0 14.889   0.045     0.047        6.9 11.617   0.034     0.035       39.5 10.166   0.029     0.030      128.8    222    000     CAA            1.787321            259.085    3.2720    1.4510    4.7230 

\begin{deluxetable}{lcccccccc}
\tabletypesize{\footnotesize}
\tablewidth{7.2truein}
\tablecaption{\textbf{X-ray Spectra of \CXOU}}
\tablehead{
\colhead{Mission/Instrument} & \colhead{ObsID} & \colhead{Flux} & \colhead{$\Gamma$} & \colhead{$N_H$} & \colhead{$\chi^2$/dof} & \colhead{$kT$} & \colhead{$N_H$} & \colhead{$\chi^2$/dof}  \\
  &   & ergs cm$^{-2}$ s$^{-1}$ &  & $10^{22}$ cm$^{-2}$ &  & keV & $10^{22}$ cm$^{-2}$ &
}
\startdata
 &  & \multicolumn{4}{c}{---------------------------Power-law fit------------------------------} & \multicolumn{3}{c}{----------Bremsstrahlung fit-----------} \\
\hline \\
\Chandra/ACIS-I & 8199 & $7^{+24}_{-3}\times10^{-13}$ & $0.5^{+2.5 a}_{-0 a}$ & $24^{+37}_{-7}$ & 1.5/3 & $>2.4$ & $29^{+36}_{-7}$ &  1.7/3 \\ %done
\Chandra/ACIS-I & 8200 &  $2.1^{+2.2}_{-1.6}\times10^{-12}$ & $3^{+0 a}_{-2.5 a}$ & $48^{+24}_{-27}$ & 147/30$^b$  & $>0.5$ & $58^{+69}_{-20}$  & 146/99$^b$ \\ % redone
\Chandra/ACIS-I & 8202 &  $1.3^{+0.2}_{-0.2}\times10^{-12}$ & $0.8^{+0.5}_{-0.3 a}$ & $1.5^{+0.8}_{-0.6}$ & 1.23/16 & $>25$ & $2.2^{+0.5}_{-0.3}$ & 1.39/16 \\ %done
\Chandra/ACIS-S & 7526 &  $1.7^{+1.0}_{-0.8}\times10^{-13}$ & $0.5^{+0.6}_{-0 a}$ & $2.8^{+4.7}_{-1.7}$ & 148/62$^b$ & $>10$ & $3.6^{+6}_{-1.8}$ & 152/85$^b$ \\ %redone
\enddata
\tablecomments{Fluxes are unabsorbed, 2-10 keV, using best-fit absorbed powerlaw model (constraining photon index to range 0.5 to 3).   \\
$^a$: Error calculation reached hard limit.\\
$^b$: Used C-statistic, reporting C-statistic value and ``goodness'' (see text) instead of $\chi^2$ and degrees of freedom.
\label{tab:cxou}}
\end{deluxetable}

\end{document}